\documentclass[%
 aip,
 amsmath,amssymb, floatfix,
 reprint,%
]{revtex4-1}

\usepackage{graphicx}% Include figure files
\usepackage{dcolumn}% Align table columns on decimal point
\usepackage{bm}% bold math
\usepackage[colorlinks=true, linkcolor=black, citecolor=black, urlcolor=black]{hyperref}
\usepackage{siunitx}
\usepackage{amsmath}
\usepackage{environ}
\usepackage[utf8]{inputenc}
\usepackage[T1]{fontenc}
\usepackage{mathptmx}
\usepackage{etoolbox}
\usepackage{float}
\usepackage{physics}

\makeatletter
\def\@email#1#2{%
 \endgroup
 \patchcmd{\titleblock@produce}
  {\frontmatter@RRAPformat}
  {\frontmatter@RRAPformat{\produce@RRAP{*#1\href{mailto:#2}{#2}}}\frontmatter@RRAPformat}
  {}{}
}%
\makeatother
\begin{document}

\preprint{AIP/123-QED}

\title[Enhancing light emission with electric fields in polar nitride semiconductors]{Enhancing light emission with electric fields in polar nitride semiconductors}
% Force line breaks with \\
\author{Nick Pant*}
\affiliation{ 
Applied Physics Program, University of Michigan, Ann Arbor, MI, USA 48109} %
\affiliation{ 
Department of Materials Science \& Engineering, University of Michigan, Ann Arbor, MI, USA 48109}%

\author{Rob Armitage}
\affiliation{ 
Lumileds, LLC, San Jose, CA, USA 95131}%

\author{Emmanouil Kioupakis}
\affiliation{ 
Department of Materials Science \& Engineering, University of Michigan, Ann Arbor, MI, USA 48109}%

\begin{abstract}
{\footnotesize Author to whom correspondence should be addressed: Nick Pant, nickpant@umich.edu \\  *Current Institution: University of Texas at Austin, Austin, TX, USA 78712}\\
\vspace{-2mm}\\
\noindent
\textbf{Abstract}\\
\noindent
Significant effort has been devoted to mitigating polarization fields in nitride LEDs, as these fields are traditionally viewed as detrimental to light emission, particularly for red emission. Contrary to this prevailing notion, we demonstrate that strong polarization fields can enhance the optical-transition strength of AlInGaN quantum wells emitting in the red, which has been historically challenging to achieve. By leveraging machine-learning surrogate models trained on multi-scale quantum-mechanical simulations, we globally explore the heterostructure design space and uncover that larger fields correlate with higher electron-hole overlap. This relation arises from the quantum-confined Stark effect, which enables thinner wells without requiring higher indium compositions, thus overcoming a key limitation in nitride epitaxy. Structural and compositional engineering of internal fields offers a unique dimension for designing polychromatic nitride LEDs, crucial for miniaturizing LED pixels to the micron scale for extended-reality and biomedical applications.  Broadly, our work demonstrates how machine learning can uncover unexpected paradigms for semiconductor design.
\end{abstract}
\keywords{micro-LED, light-emitting diode, quantum well, III-nitride, polarization fields, machine learning, multi-scale modeling, red emission}

\maketitle

\section{Introduction}

Internal electric fields arise from polarization discontinuities at the interfaces of  semiconductors with broken inversion symmetry. Such fields have been used to induce free carriers in wide-bandgap semiconductors without dopant impurities,\cite{chaudhuri2019polarization} to enable switchable non-volatile memory,\cite{kim2023wurtzite} and to separate electrons from holes to improve the performance of solar cells and photocatalysts.\cite{yuan2011efficiency, frost2014atomistic, kibria2016atomic}  However, polarization has traditionally been considered detrimental for light-emission applications, such as light-emitting diodes (LED) and lasers, where efficient electron-hole recombination is essential.\cite{fiorentiniEffectsMacroscopicPolarization1999, Waltereit2000, pimputkarProspectsLEDLighting2009, dreyerCorrectImplementationPolarization2016, aufdermaurEfficiencyDropGreen2016, lin2023micro}

Nevertheless, wurtzite InGaN/GaN quantum wells, featuring strong spontaneous and piezoelectric polarization fields ($>$1 MV/cm), underpin commercial solid-state lighting.\cite{pimputkarProspectsLEDLighting2009} Because the band gap of these LEDs can be tuned across the visible spectrum, they are ideally suited for integrating blue, green, and red wavelengths onto a monolithic platform, crucial for creating miniaturized polychromatic pixels that have applications in extended reality and biomedicine.\cite{lin2020development, behrman2022} However, improving the energy efficiency of InGaN for red emission remains challenging.\cite{lin2023micro}

\begin{figure*}[ht!]
    \centering
    \includegraphics{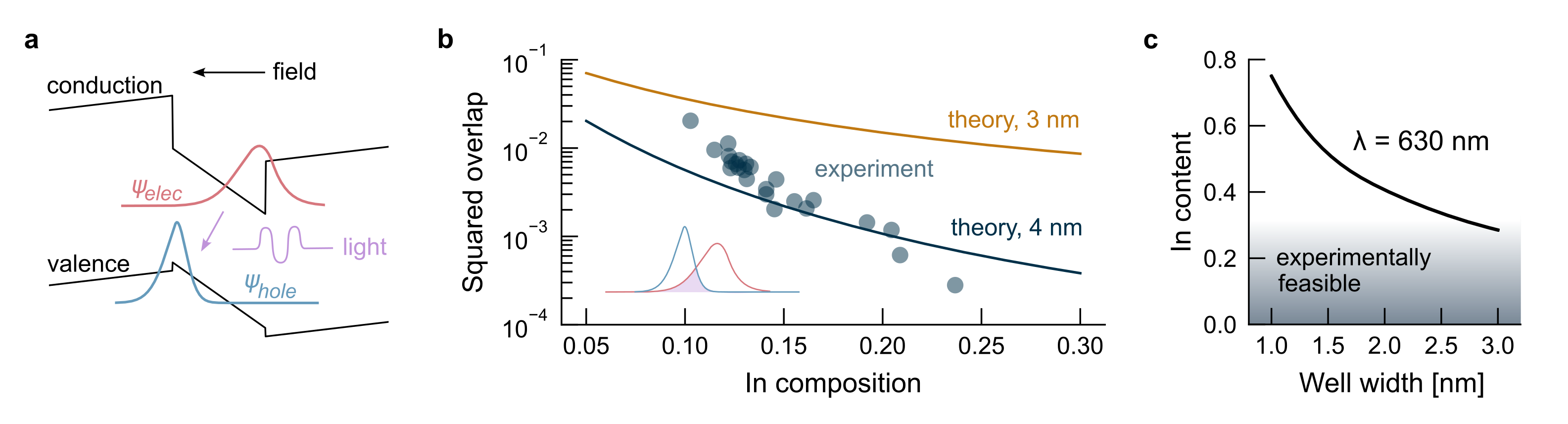}
    \caption{\footnotesize {Origin of the red efficiency gap in InGaN LEDs. \textbf{a,}} Band diagram of InGaN quantum wells, showing how internal polarization fields separate electron and hole wave functions, reducing their overlap and probability of recombination. {\small \textbf{b,}} An increase in the In composition increases the polarization field, which dramatically decreases the wave-function overlap. The square of the overlap is correlated with the $B$ coefficient of quantum wells, $B_\text{QW}$, relative to their bulk value, $B_\text{bulk}$; experimentally measured $B_\text{QW}/B_\text{bulk}$ in InGaN/GaN single quantum wells\cite{davidQuantumEfficiencyIIINitride2019} are shown for qualitative comparison. {\small \textbf{c,}} To improve the wave-function overlap within quantum wells, the simplest strategy is to decrease the well width, however this is challenging in conventional InGaN/GaN quantum-wells as thinner wells require In compositions higher than thermodynamically favored, in order to maintain red emission.}
    \label{fig:problem}
\end{figure*}

The drop in efficiency of InGaN LEDs with increasing wavelength is attributed to the field-induced separation of electron and hole wave functions in their quantum wells [Fig. \ref{fig:problem}a].\cite{aufdermaurEfficiencyDropGreen2016, lin2023micro} This reduces the optical-transition dipole moment, which is proportional to the wave-function overlap. Achieving red emission requires high In composition, which increases internal electric fields and reduces radiative recombination, quantified by the $B$ coefficient [Fig. \ref{fig:problem}b]. Unfortunately, making the quantum wells thinner, to bring electrons and holes closer, shifts the emission to shorter wavelengths, necessitating even higher In compositions to maintain red emission [Fig. \ref{fig:problem}c].

The challenges associated with achieving high-quality InGaN with over 30\% In have been extensively discussed,\cite{schulz2020influence, mishraUnlockingOriginCompositional2021, huUnderstandingIndiumIncorporation2022} and is connected to a surface reconstruction during growth that limits In incorporation.\cite{lymperakisElasticallyFrustratedRehybridization2018a}  Mitigating the polarization of nitride LEDs\cite{Waltereit2000, pimputkarProspectsLEDLighting2009, youngPolarizationFieldScreening2016, lin2023micro} has yielded limited success at long wavelength because non-polar designs need more than 40\% In for red emission.

Here, we identify red-emitting quantum wells with the highest electron-hole wave-function overlaps, by systematically exploring the configuration space of Al$_x$In$_{1-x-y}$Ga$_{y}$N quantum wells, restricting the search to realistic compositions of 30\% or less In. By taking a panoramic view, we aim to identify overarching \textit{design principles} that are difficult to access with laborious experiments limited to exploring a small region of the design space. We addressed the combinatorial challenge of such an exploration through ensemble machine-learning (ML) algorithms that provide a 340$\times$ speedup over quantum-mechanical calculations, reproducing a wide range of experimental measurements\cite{davidQuantumEfficiencyIIINitride2019, xue2023recombination} without any adjustable parameter.

To our surprise, increasing the electric field increases the wave-function overlap in red-emitting quantum wells, as the quantum-confined Stark effect (QCSE) enables a reduction in well width while keeping the emission wavelength constant. The wave-function overlaps of the identified red emitters are comparable to state-of-the-art blue emitters. We \textcolor{black}{propose} that an enhancement of electric fields could partially explain why incorporating Al into the barrier or relaxing strain within the quantum well improves the performance of state-of-the-art red LEDs. \textcolor{black}{These strategies have been shown to improve LED performance in several experimental works,~\cite{hwangDevelopmentInGaNbasedRed2014, ozaki2018red, iida2020high, LiSignificantQuantumEfficiency2023} though the beneficial role of polarization fields was not previously identified.}

\section{Machine-learning accelerated search}

ML surrogate models have emerged as a promising approach for accelerating expensive quantum-chemical calculations to virtually screen organic molecules\cite{gomez2016design} and inorganic compounds.\cite{choubisa2023interpretable, merchant2023scaling} Here, we use ML to accelerate our exploration of the design space of AlInGaN heterostructures, which is challenging because of the combinatorial explosion of the design space with an increasing number of design variables. We trained our ML models on effective-mass Hamiltonians\cite{kohn1957effective} derived from first-principles DFT calculations, which we have previously shown can reliably model the spectral and recombination properties of blue- and green-emitting InGaN quantum wells grown at \texttt{Lumileds}.\cite{pantOriginInjectiondependentEmission2022, liCarrierDynamicsBlue2023} See Methods in Section 1 of the Supporting Information (SI) for how we built the Hamiltonian, solved Schr\"{o}dinger's and Poisson's equations, variationally minimized the exciton problem, and tested the speedup afforded by the ML algorithm. 

\begin{table}[h!]
\fontfamily{ppl}\selectfont\footnotesize
\centering
\begin{tabular}{l|ll|ll}
\textbf{Configuration} & \multicolumn{2}{l|}{\textbf{Emission [eV]}} & \multicolumn{2}{l}{\textbf{Squared overlap}} \\
\textbf{} & Theory & Expt. & Theory & $B_\text{QW}^\text{expt.}/B_\text{bulk}$ \\ \hline
\begin{tabular}[c]{@{}l@{}}3 nm In$_{0.3}$Ga$_{0.7}$N\\ 18 nm GaN\end{tabular} & 1.90 & N/A & 0.0069 & N/A \\ \hline
\begin{tabular}[c]{@{}l@{}}3.1 nm In$_{0.29}$Ga$_{0.71}$N\\ 1.2 nm Al$_{0.90}$Ga$_{0.10}$N\\ 9.6 nm GaN\end{tabular} & 1.91 & 1.90 & 0.0083 & 0.0051 \\ \hline
\begin{tabular}[c]{@{}l@{}}3.3 nm In$_{0.27}$Ga$_{0.73}$N\\ 1.2 nm Al$_{0.82}$Ga$_{0.18}$N\\ 9.5 nm GaN\end{tabular} & 1.97 & 1.92 & 0.0056 & 0.0030 \\ \hline
\begin{tabular}[c]{@{}l@{}}3.5 nm In$_{0.28}$Ga$_{0.72}$N\\ 1.25 nm Al$_{0.82}$Ga$_{0.18}$N\\ 9.0 nm GaN\end{tabular} & 1.85 & 1.92 & 0.0030 & 0.0022 \\ \hline
\begin{tabular}[c]{@{}l@{}}3.7 nm In$_{0.28}$Ga$_{0.72}$N\\ 1.2 nm Al$_{0.82}$Ga$_{0.18}$N\\ 10.4 nm GaN\end{tabular} & 1.79 & 1.89 & 0.0016 & 0.0011 \\ \hline
\end{tabular}%
%}
\caption{Benchmark of quantum calculations against experiments. Our quantum-mechanical calculations accurately predict the salient features of red-emitting quantum wells reported in the literature,\cite{xue2023recombination} without any adjustable parameter. Our ML models are trained on data generated by these calculations. }
\label{tab:my-table}
\end{table}

As a starting point, we model ideal quantum wells without alloy disorder or well-width fluctuations. By considering these effects, we have verified that they do not change our main conclusions in Sections 3 and 6 of the SI.
 
In Table I, we benchmark our quantum-mechanical calculations against experimental measurements of red InGaN emitters from the literature.\cite{xue2023recombination} It is notable that our idealized model predicts the experimental photoluminescence energies with a mean absolute error of 0.06 eV. Investigating the remaining discrepancy in the prediction of the emission energy requires a careful analysis of the experimental structures measured by Xue et al.\cite{xue2023recombination} While it is important to recognize the limitations of our model in predicting emission energies of \textit{specific} LEDs, for which microscopic structural details matter, an accuracy within 0.1 eV is nonetheless sufficient to identify \textit{global} design trends. 
We were unable to validate our predictions on thinner quantum wells, for which experimental characterization data remain scarce. Thinner quantum wells are expected to feature stronger excitonic effects, which we have accounted for in our emission predictions with a variational approach (see Methods in Section 1 of the SI).

Although the squared overlap of the lowest eigenstates is not directly measurable, we show that it exactly correlates with the decrease in the radiative recombination $B$ coefficient of quantum wells, $B_\text{QW}$, relative to the bulk value of 6$\times10^{-11}$ cm$^3$/s.\cite{kioupakis2013temperature, schiavonWavelengthdependentDeterminationRecombination2013} The calculated squared overlaps systematically predict the trends in experimental $B_\text{QW}/B_\text{bulk}$. Additional benchmarks of our calculations against experimentally measured $B$ coefficients\cite{davidQuantumEfficiencyIIINitride2019} at different wavelengths are provided in Fig. \ref{fig:problem}b for InGaN/GaN single quantum wells.

We defined our search space using practical considerations. We considered quantum wells separated from their periodic images by large barriers of 18 nm to, initially, focus on the physics of single quantum wells. We employed $c$-plane GaN substrates reflecting the current standard for LED epitaxy, and considered well widths between 0.3 nm and 4.0 nm, as thicker wells are likely to relax strain by generating defects. Lastly, we required the In composition in both the well and barrier to be 30\% or less.

\begin{figure*}[htp!]
    \centering
    \includegraphics{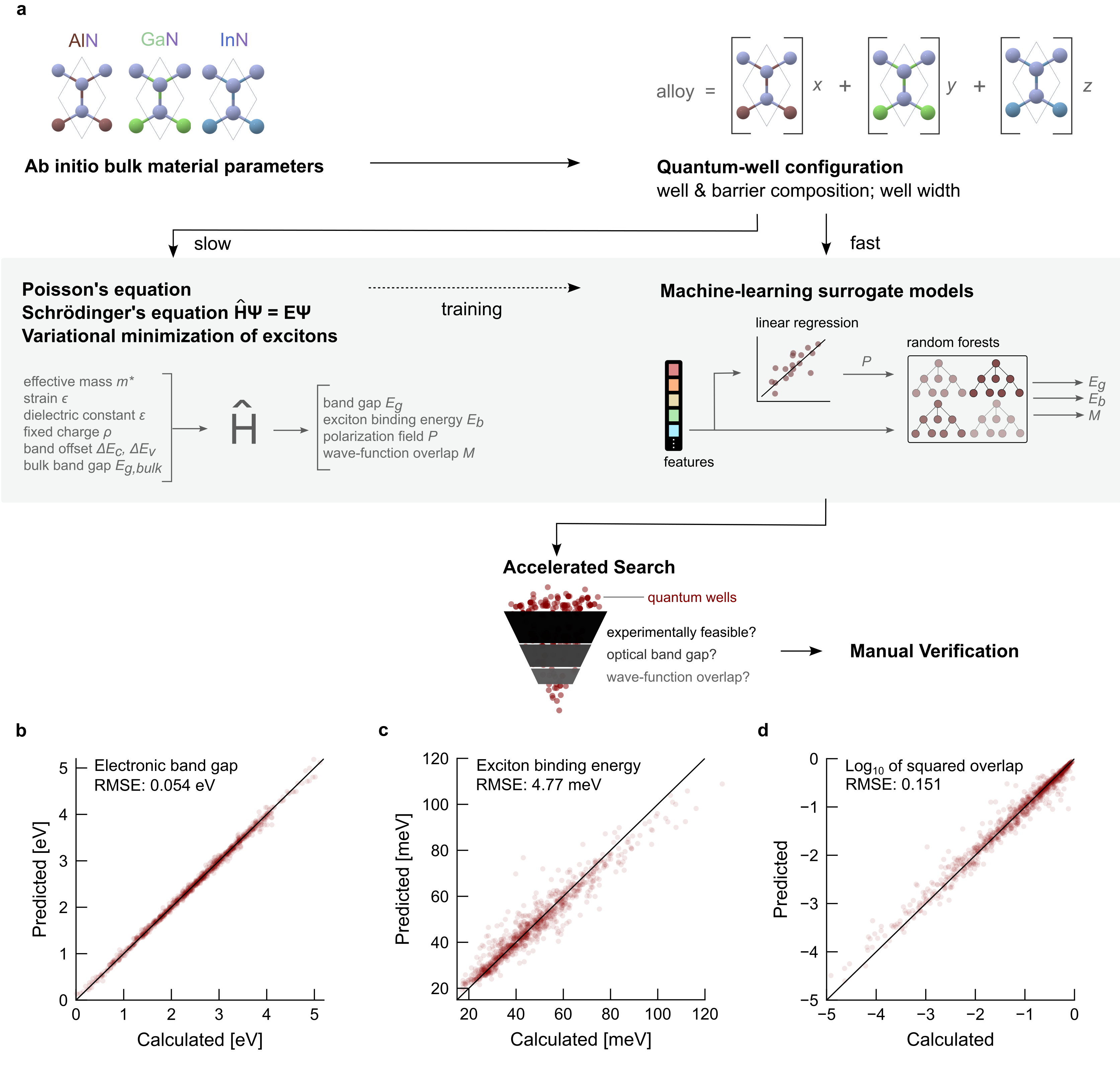}
    \caption{ Machine learning for heterostructure exploration. {\small\textbf{a,}} Diagram of the ML accelerated workflow: we employed ensemble ML models, trained on first-principles-derived quantum-mechanical data, as surrogates for efficiently screening candidate quantum-well configurations. We filtered the candidates according to the experimental feasibility of their composition, the optical band gap, and the wave-function squared overlap. {\small\textbf{b,}} Performance of the ML surrogates for predicting the electronic band gap compared to quantum-mechanical calculations. {\small\textbf{c,}} The ML predicted exciton binding energy compared to quantum-mechanical calculations. {\small\textbf{d,}} The ML predicted squared overlap of electron and hole wave functions compared to quantum-mechanical calculations. }
    \label{fig:workflow}
\end{figure*}

We illustrate our ML accelerated workflow in Fig. \ref{fig:workflow}a. Initially, a linear-regression model regularized with the least absolute shrinkage and selection operator (LASSO) predicts polarization fields within quantum wells. Subsequently, these predicted fields are cascaded to random decision forests, which predict the target variables by learning non-linear relationships using ensembles of decision trees. We chose these models because they are widely available, computationally inexpensive, and easily trained on outputs of any numerical solver. We provide details in Methods. Fig. \ref{fig:workflow}b,c,d shows predicted vs quantum-mechanically calculated electronic band gap, exciton binding energy, and base-10 logarithm of the wave-function squared overlap. These benchmarks are calculated on a hold-out dataset of 4000 randomly sampled configurations in the entire configuration space, not restricted to In $\leq$ 0.3. 

\begin{figure*}[htp]
    \centering
    \includegraphics{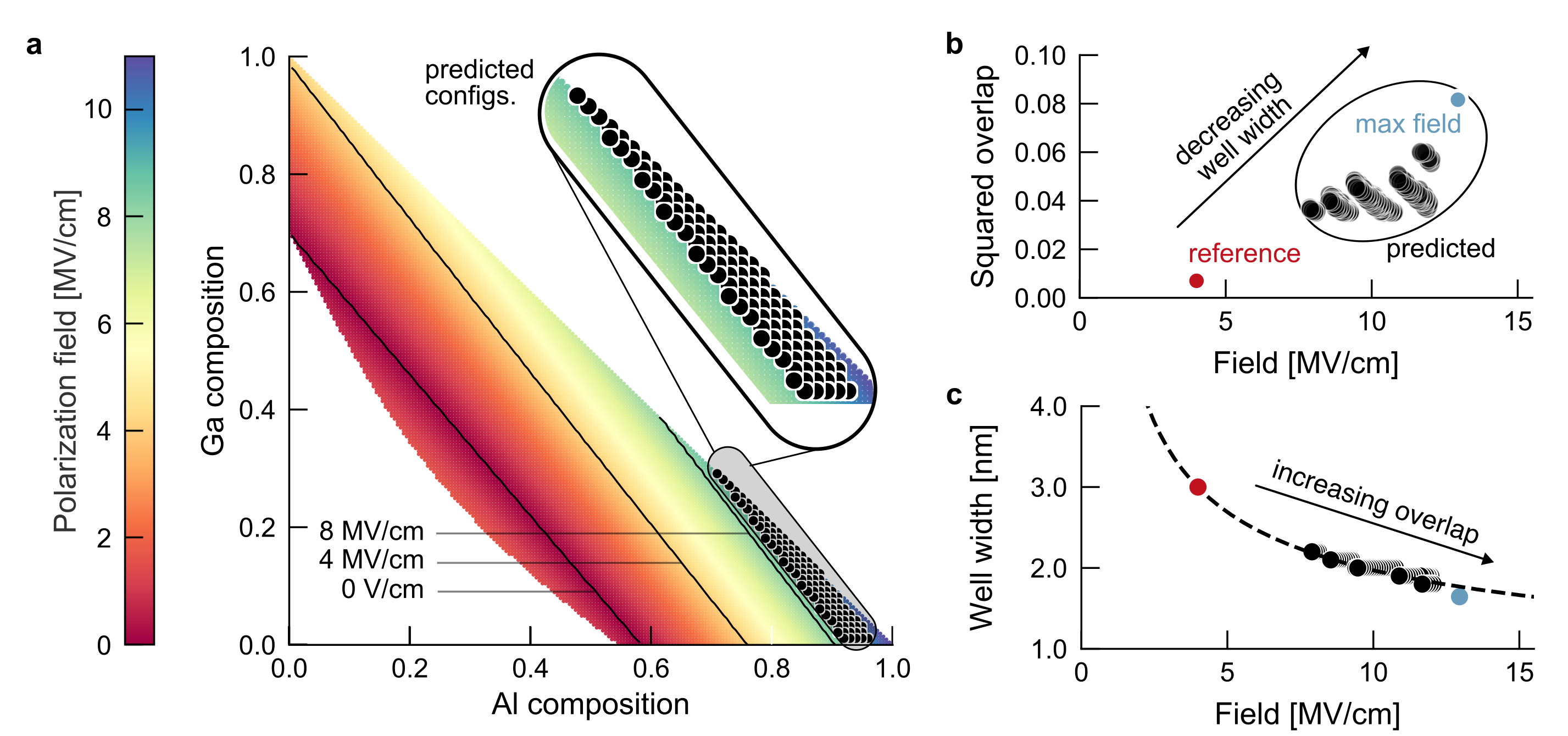}
    \caption{  Role of extreme fields in the optimal red-emitting quantum wells. {\small \textbf{a,}} Map of the polarization field across an In$_{0.3}$Ga$_{0.7}$N well as a function of the AlInGaN barrier composition. The predicted configurations (black points) cluster in a narrow region of configuration space, bounded by contours corresponding to extremely large fields. {\small \textbf{b,}} Despite exhibiting significantly larger fields, the predicted quantum wells (black and blue points) exhibit greater wave-function overlap than the reference InGaN/GaN quantum well (red point). ``Max field" refers to the In$_{0.3}$Ga$_{0.7}$N/AlN configuration, which maximizes the field within our design space. {\small \textbf{c,}} Dependence of well width on field strength for In$_{0.3}$Ga$_{0.7}$N wells with an optical band gap of 1.82 eV. The dashed curve is a guide to the eye. Increasing the electric field enables the well width to decrease, improving the optical oscillator strength.}
    \label{fig:statistics}
\end{figure*}

Using the trained surrogate models, we exhaustively searched the configuration space, screening for optical band gaps between 1.775 eV (698 nm) and 1.925 eV black{(644 nm)}, which corresponds to red emission. Initially, we screened for large oscillator strengths by filtering for wave-function squared overlaps that are at least 3$\times$ larger than reference 3 nm wide In$_{0.3}$Ga$_{0.7}$N/GaN well. \textcolor{black}{Manual} verification with \textcolor{black}{explicit} quantum-mechanical calculations yielded 6591 candidates in total. Out of this pool, 851 candidates feature wave-function squared overlaps at least 5$\times$ larger than the reference; we hereafter refer to these 851 designs as "optimal." 

\section{Optimal Designs}

We present a survey of the the optimal quantum wells in Fig. S1. Interestingly, every optimal well is between 1.8 nm and 2.2 nm thick, as thinner wells require In composition greater than 30\% to reach 1.85 eV emission and thicker wells are unable to reach $5\times$ greater squared wave-function overlap than the reference InGaN/GaN quantum well. Moreover, the wells uniformly contain between 26\% to 30\% In, with minimal to no Al, while the barriers are rich in Al and sometimes also contain In.  Further configurational details can be found in Supporting Dataset 1.

The most striking feature of the optimal configurations is their extremely large polarization fields, ranging from 8 MV/cm to 12 MV/cm. These fields are related to the polarization mismatch between the well and barrier materials. The relationship between barrier composition and polarization fields within the well is best elucidated through a contour map, shown in Fig. \ref{fig:statistics}a, where we approximate the well material as In$_{0.3}$Ga$_{0.7}$N since the optimal designs have nearly identical well compositions. Upon superimposing upon this map the barrier compositions of the optimal configurations, it becomes evident that the configurations cluster within a narrow region defined by the contour lines of extremely large polarization fields. We have shown the coordinates of the 2.0 nm-wide optimal quantum wells in Fig. \ref{fig:statistics}a to streamline the illustration, but this analysis applies to other well widths as well. 

Despite their extreme fields, the wave-function overlaps in the optimal wells surpass those of the reference well with GaN barriers, as illustrated in Fig. \ref{fig:statistics}b. Our analysis also suggests that the optimal quantum wells exhibit $B$ coefficients around $\sim 10^{-11}$ cm$^3$/s, comparable to state-of-the-art blue emitters.\cite{liCarrierDynamicsBlue2023} Although the Al-rich barriers confine carriers more strongly within the well because of their larger band offsets,\cite{vichi2020increasing, xue2023recombination} it is the polarization field rather than improved carrier confinement that causes the primary improvement in wave-function overlap (Fig. \ref{fig:statistics}c). We discuss our numerical control experiments to probe this hypothesis in Section 4 of the SI and Fig. S2. 

As the final verification, we push this concept to its logical extreme by considering In$_{0.3}$Ga$_{0.7}$N wells with AlN barriers, a combination that maximizes the internal field within our design space. Detailed calculations reveal a field strength of 13 MV/cm and an optical band gap of 1.82 eV at a well width of 1.62 nm. The wave-function overlap associated with this configuration is indeed an order of magnitude larger than that of 3 nm InGaN/GaN quantum wells, as shown in Fig. \ref{fig:statistics}b. Our initial search space did not encompass well widths of 1.62 nm, reflecting the practical necessity of discretizing continuous variables to numerically represent them. Nevertheless, this limitation did not hinder us from identifying the broader design rule that ultimately led to the discovery of this configuration. \textcolor{black}{While it is not presently feasible to use pure AlN barriers in actual LEDs, this example illustrates the fact that polarization fields need not be mitigated to enhance the wave-function overlap.}

In addition to compositional engineering, there is more freedom in modifying polarization fields by exploring qualitatively different heterostructure designs \textcolor{black}{(explored in Section 5 of the SI)}. We facilitated this exploration by considering InGaN quantum wells separated from GaN barriers with AlInGaN interlayers, a class of heterostructures already employed in state-of-the-art LEDs.\cite{hwangDevelopmentInGaNbasedRed2014, iida2020high, LiSignificantQuantumEfficiency2023} We recover the limit of quantum wells with pure AlInGaN barriers, corresponding to our initial search space, by setting the interlayer thickness equal to the barrier thickness. To streamline our understanding, we focused on AlN interlayers, although our conclusions apply to other compositions. 

To calculate a geometric map of polarization fields, we solved Poisson's equation in configurations with 2 nm wide In$_{0.3}$Ga$_{0.7}$N wells, and we systematically varied the thickness of the interlayers and the barriers, considering interlayers placed on one or both sides of the wells. We modeled structures that are coherently strained on GaN and also considered partially and fully strain-relaxed wells, reflecting emerging schemes that attempt to improve material quality by relaxing strain.\cite{zhang2020efficient, dussaigneFullInGaNRed2021, ji2023porous, pandey2023red} The resulting maps of polarization fields are illustrated in Fig. S3 and S4.

Engineering the heterostructure geometry and relaxing their strain enables the precise control of internal fields. For example, the highest polarization fields correspond to symmetric barriers, which maximize the mismatch in polarization on both sides of the well. Additionally, larger interlayer thicknesses tend to increase the field across the well; increasing the interlayer and barrier thickness simultaneously has the largest impact on amplifying fields. In contrast, making the barrier thicker without also making the interlayer thicker \textit{decreases} the polarization field. 

Moreover, we find that engineering strain can also modify the fields. For example, relaxing strain within the well can advantageously enhance the fields because it increases the polarization mismatch between the barrier/interlayer and well regions. Therefore, elastic strain relaxation can be used to redshift the band gap and further decrease the well width to promote wave-function overlap. Examples include the incorporation of In into the substrate to minimize the lattice mismatch with the active region \cite{evenEnhancedIncorporationFull2017} and the electrochemical etching of the substrate to modify its elastic response.\cite{ji2023porous} These geometric design rules can be combined with compositional engineering to enhance the wave-function overlaps in quantum wells. 

\section{Discussion}

Polarization fields have been known to redshift InGaN's band gap and prior works have attempted to leverage this effect to improve the performance of LEDs.\cite{vichi2020increasing, koleske2015increased} However, this has received limited attention because polarization has been fundamentally viewed as a factor that reduces wave-function overlap. To the contrary, by globally mapping the design space of feasible AlInGaN heterostructures, we have demonstrated that designs with the largest wave-function overlaps exhibit large electric fields because they support thinner wells.

While it is true that non-polar quantum wells exhibit unity wave-function overlap, they require infeasibly high In concentrations exceeding 40\% and well widths substantially thicker than 3 nm for red emission. Promising approaches are emerging that improve In incorporation through strain relaxation,\cite{zhang2020efficient, dussaigneFullInGaNRed2021, ji2023porous, pandey2023red} however no planar LED with such high In content and wide wells have ever been reported with associated external quantum efficiency greater than 1\%. 

Notably, our findings align with experimental reports of the most efficient red LEDs, many of which feature Al-containing barriers or strain-relaxed quantum wells.\cite{hwangDevelopmentInGaNbasedRed2014, iida2020high, LiSignificantQuantumEfficiency2023}  Hence, we suggest critically reevaluating whether polarization fields already play a role in improving the performance of these LEDs. Overall, carriers recombine more quickly in quantum wells with higher wave-function overlap, meaning fewer carriers are needed to achieve high current densities. This helps to reduce the carrier-density-dependent hue shift\cite{pantOriginInjectiondependentEmission2022} and mitigate efficiency droop, a phenomenon where the efficiency decreases with increasing carrier density.\cite{liCarrierDynamicsBlue2023} \textcolor{black}{We underscore that the most extreme fields are not needed to reap the benefits. Even slight increases in the fields allow quantum wells to be thinner without increasing the In content, thanks to the QCSE.} 

However, it remains an open question if further enhancing polarization fields will lead to substantial benefits when considering other aspects of LED performance. Notably, we have not explicitly modeled well-width fluctuations, inter-well carrier transport, and device-level characteristics, all of which present ongoing challenges for predictive modeling. Furthermore, Al-containing barriers and large electric fields in LEDs introduce several engineering hurdles. These include the competition between radiative and Auger-Meitner recombination, spectral linewidth broadening, \textcolor{black}{material challenges} during growth, and challenges with electrical injection. We discuss these issues and potential solutions in the SI Section 7, \textcolor{black}{including identifying nearly lattice-matched designs provided in Supporting Dataset 2. We also refer the reader to Ref.~\citenum{li2025impactquantumthicknessefficiency} in which we discuss the challenges of thin quantum wells in the context of green LEDs.}

\section{Conclusion}

III-nitrides are a cornerstone of optoelectronics. With their distinct crystal structure in the III-V family and their proliferation in industry, they offer an unmatched platform to harness polarization effects. While we do not claim to solve the challenges associated with \textcolor{black}{\textit{further}} amplifying electric fields, we note that designs mitigating these fields face comparable, if not greater, challenges. This is a salient point to consider as the community considers moving towards nonpolar designs.\cite{lin2023micro} The ability to exploit the unique polarization degree of freedom in the design of optoelectronics should not be underestimated, especially with the addition of new elements to the III-nitride family\cite{jena2019new} that introduce ferroelectric\textcolor{black}{ity}.

We have shown that internal fields do not need to be mitigated to enhance wave-function overlap. This is compelling given that polarization fields are the reason why thin InGaN quantum wells ($\leq 3$ nm) with low In concentrations ($\leq$ 30\%), used in red LEDs, can emit red light in the first place. \textcolor{black}{Despite two decades of efforts to mitigate polarization fields, the most efficient visible LEDs, especially in Industry, exhibit fields on the order of MV/cm. Hence, polarization is a double-edged sword, which, despite its drawbacks, also benefits LEDs through the QCSE by enabling long wavelength emission in thin wells with low In concentrations.} \\

\section*{Associated Content}
\textbf{Supporting Information.} 
Additional details of the computational methods and parameters used (PDF). Survey of the optimal quantum wells and geometric blueprint for extreme fields (PDF). Discussion of impact of alloy fluctuations, carrier confinement, emission broadening due to well-width and alloy fluctuations, challenges associated with large electric fields, and electron overflow (PDF). Dataset of ideal quantum-well configurations (Dataset 1) and nearly lattice-matched configurations (Dataset~2).\\

\section*{Acknowledgements}
N.P. thanks Kyle Bushick, Mahlet Molla, Emily Oliphant, and Amanda Wang for their feedback on the manuscript.

\textbf{Funding Sources.}
This project was funded by the U.S. Department of Energy, Office of Energy Efficiency and Renewable Energy, under Award No. DE-EE0009163. Computational resources were provided by the National Energy Research Scientific Computing Center, a Department of Energy Office of Science User Facility, supported under Contract No. DEAC0205CH11231. N.P. acknowledges the support of the Natural Sciences \& Engineering Research Council of Canada Postgraduate Scholarship.\\

% \section*{References}
\textbf{References.}
\bibliography{ref}% Produces the bibliography via BibTeX.

\end{document}